\documentclass[useAMS,usenatbib]{mn2e}

\usepackage{natbib, aas_macros}
\usepackage{ulem}
\usepackage[dvips]{graphicx}
\usepackage{wrapfig}
\graphicspath{{./figs/}}
\usepackage{color}
\usepackage{amsmath}
\usepackage{amssymb}

\newcommand{\Msun}{{\rm  M_{\odot}}}

\newcommand{\La}{L_{\rm Ly\alpha}}

\newcommand{\Mstar}{M_{\rm{star}}}

\newcommand{\lya}{\rm {Ly{\alpha}}}

\newcommand{\Mh}{M_{\rm h}}
\newcommand{\Msunyr}{\rm M_{\odot}~ yr^{-1}}
\newcommand{\fescion}{f_{\rm esc}^{\rm ion}}

\newcommand{\Rvir}{R_{\rm vir}}

\newcommand{\Fdrag}{F_{\rm drag}}

\newcommand{\rhoa}{\rho_{\rm Ly\alpha}}
\newcommand{\Na}{N_{\rm Ly\alpha}}
\newcommand{\ea}{\epsilon_{\rm Ly\alpha}}
\newcommand{\ft}{f_{\rm trap,15}}
\newcommand{\tdyn}{t_{\rm dyn}}
\newcommand{\tang}{t_{\rm ang}}
\newcommand{\Mcrit}{M_{\rm crit}}
\newcommand{\Msf}{\dot{M}_{\rm *}}
\newcommand{\fgas}{f_{\rm gas}}
\newcommand{\Voutmax}{V_{\rm out, max}}
\newcommand{\esf}{ \epsilon_{\rm SF}}
\newcommand{\Tvir}{T_{\rm vir}}
\newcommand{\fang}{f_{\rm ang}}

\title[The mass and luminosity of POPIII galaxies]
{
Upper limits on the mass and luminosity of Population III-dominated galaxies \\
}
\author[Yajima et al.]
{Hidenobu Yajima$^{1, 2}$\thanks{E-mail: yajima@astr.tohoku.ac.jp (HY)} and Sadegh Khochfar$^{3}$ 
\\
$^{1}$ Frontier Research Institute for Interdisciplinary Sciences, Tohoku University, Sendai 980-8578, Japan\\
$^{2}$ Astronomical Institute, Tohoku University, Sendai 980-8578, Japan\\
$^{3}$ 
Institute for Astronomy, University of Edinburgh, Royal Observatory, Edinburgh, EH9 3HJ, UK\\
}
 
\begin{document}

\date{Accepted ?; Received ??; in original form ???}

\pagerange{\pageref{firstpage}--\pageref{lastpage}} \pubyear{2008}

\maketitle

\label{firstpage}

%
%
\begin{abstract}
We here derive upper limits on the mass and luminosity of  Population III (POPIII) dominated proto-galaxies based on the collapse of primordial gas under the effect of angular momentum loss via $\lya$ radiation drag and the gas accretion onto a galactic centre. 
Our model predicts that POPIII-dominated galaxies at $z \sim 7$ are hosted by haloes with $\Mh \sim 1.5 \times 10^{8} - 1.1 \times 10^{9}~\Msun$, 
that they have $\lya$ luminosities of $\La \sim 3.0 \times10^{42} - 2.1 \times 10^{43}~\rm erg~s^{-1}$, stellar mass of $\Mstar \sim 0.8 \times 10^{5} - 2.5 \times 10^{6}~\Msun$, and outflowing gas with velocities $V_{\rm out} \sim 40~\rm km~s^{-1}$ due to $\lya$ radiation pressure.
We show that the POPIII galaxy candidate CR7 violates the derived limits on stellar mass and   $\lya$ luminosity and thus is unlikely to be a POPIII galaxy.
 POPIII-dominated galaxies at $z \sim 7$ have He{\sc ii} line emission that is $ \sim1- 3$ orders of magnitude lower then that of $\lya$,  they have high $\lya$ equivalent width of $\gtrsim 300~\rm \AA$  and should be found close to bright star forming galaxies. The  He{\sc ii} $1640~\rm \AA$ line is in comfortable reach of  next generation telescopes, like the  JWST or TMT. 
\end{abstract}

%
%

%
%
\section{Introduction}
The current paradigm of structure formation predicts the formation of the first generation of stars, so-called Population III (POPIII) stars, in mini-haloes with mass $\Mh \sim 10^{6}~\Msun$ at $z \gtrsim 15$ 
\citep[e.g.,][]{Abel02, Bromm02, Omukai03, Yoshida08, Stacy14, Susa14}.
Following the initial POPIII episode metal pollution from POPIII supernovae pollute the surrounding inter-stellar medium and trigger the transition to POPII star formation \citep{Maio11, Johnson13, Jeon15}.
Recent cosmological simulations indicated that  POPII star formation starts dominating over POPIII star formation at $z \lesssim 10$ 
\citep[e.g,][]{Johnson13}.
However, POPIII stars are able to form locally due to inhomogeneous metal enrichment \citep{Tornatore07},
and suppression of star formation by external UV feedback \citep{Hasegawa13}. 
In addition, the initial mass function of POPIII stars is still under the debate \citep[e.g,][]{Hirano14, Stacy16}. 
hence the total metal production and efficiency of the enrichment process are unclear.
Understanding how stellar population transit from POPIII to POPII stars, 
and whether massive POPIII star clusters are able to form at $z \ll 10$ is crucial given the capabilities of future mission such as e.g. the {\it James-Webb-Space Telescope} (JWST) to probe the high-redshift Universe. 

If the majority of stars in a galaxy are POPIII, the galaxies will show He{\sc ii}  and  $\lya$ emission lines. The former is a distinct result of the high surface temperature of massive POPIII stars which emit photons that double ionize helium \citep[e.g.,][]{Bromm01a}. In the following we will refer to such systems as dual emitters.
\citet{Nagao08} conducted a search for  dual emitters, 
by targeting $\lya$ emitting galaxies (LAEs) with high $\lya$ equivalent width (EW). 
Because the $\lya$ EW is proportional to the ratio of ionizing photons to non-ionizing UV continuum photons \citep[see however,][]{Yajima14a, Yajima15a}, it depends on the stellar population and it is $\lesssim 300~\rm \AA$ for a Salpeter-like IMF. Hence, LAEs with unusual high EW are prime candidates to find dual emitters. 
However, the survey by \citet{Nagao08} did not detect He{\sc ii} from any LAEs in their sample, which indicates an upper limit for the cosmic star formation density of POPIII stars, $< 5 \times 10^{-6} ~\rm \Msun~yr^{-1}~Mpc^{-3}$ at $z \sim 4$. 
Recently \citet{Sobral15} have reported the detection of He{\sc ii} from a bright LAE at $z=6.6$, which is called CR7.
CR7 is very luminous in $\lya$ with $L_{\rm lya} = 10^{43.9}~\rm erg~s^{-1}$, 
has a high EW $\gtrsim 200 \rm \AA$, and no detection of metal lines (see however, Bowler et al. 2016 for metal line detections in CR7). In, addition, it is close to a system of star forming galaxies at a projected distance  of $\sim 5~\rm kpc$.
Based on SED modelling the authors suggest that CR7 could host a POPIII star cluster with $M_{*} \sim 10^{7}~\Msun$. 

In order to form  such a POPIII star cluster, a massive primordial cloud has to collapse. 
\citet{Visbal16} suggest that the collapse of CR7 could be  suppressed by photo-ionization from the neighboring galaxies, and that only once the halo grows massive enough the gas can collapse. 
\citet{Pallottini15} show that indeed cosmological hydrodynamics simulation, produce conditions as required for the formation of POPIII clusters. However, the mass they find is smaller than required by the observations of \citet{Sobral15}.    
 One main obstacle in forming a massive POPIII star clusters is that over the life-time of massive stars, a few Myrs,  the gas needs to collapse and the formation of the POPIII star cluster needs to take place. If the gas has sufficient specific angular momentum it will settle in a large scale disc and collapse will take place on a viscous time scale \citep[e.g.,][]{Ceverino10}.
 An efficient mechanism to fuel a central star burst on shorter time scales and to destroy such disc is via mergers \citep{DOnghia06}. 
However, time scales for mergers and the dropping merger rate at high redshifts limit the efficiency of this process \citep[e.g.,][]{Khochfar01}. 
 Alternatively, gas can loose its angular momentum via $\lya$-photon-drag already before settling in a disc while orbiting through a high intensity, homogenous background of $\lya$ photons \citep{Yajima14d}.
 
While $\lya-$photon drag is able to remove angular momentum, \citet{Smith16} show that intense $\lya$ flux can cause galactic outflow as well. Their 1-D $\lya$ radiative-hydrodynamics simulations show that such outflows can produce asymmetric $\lya$ line profiles with red wing which was consistent with the observations.

In this paper, we will investigate the formation of massive POPIII-dominated galaxies
by focusing on the effect that $\lya$ photons have on angular momentum transport and outflows
We will present the range of  halo masses and required $\lya$ luminosities to form massive POPIII galaxies.



%
%
\section{Model}
\label{sec:model}


%
%

\subsection{Critical distance for ionization from a star-forming galaxy}
The enrichment process of the ISM by POPIII stars is very quick and of the order $10^7$ yr \citep[e.g.,][]{Maio11}. 
This in effect allows for only one generation of POPIII stars being born in a halo \citep[e.g.,][]{Johnson13}. To form a massive POPIII star cluster thus requires a massive primordial gas cloud that can fragment and form POPIII stars in one go.
One way to achieve this is by photo-ionization heating \citep{Johnson14, Visbal16}. 
The mean UV background (UVB) is not sufficiently high enough to ionize haloes at $z > 7$  \citep{Faucher09}. 
However,  the UV flux from neighbouring star-forming galaxies can provide enough UV photons.
The condition to ionize haloes just after virialization is, 
\begin{equation}
\fescion \dot{N}_{\rm ion} \frac{\pi \Rvir^{2}}{4 \pi D^{2}} > \frac{4 \pi \Rvir^{3} \alpha_{\rm B} n_{\rm H}^{2}}{3} 
+ \pi \Rvir^{2} D \alpha_{\rm B} n_{\rm H, IGM}^{2}, 
\end{equation}
$\dot{N}_{\rm ion}$ is the ionizing photon emissivity of the star-forming galaxy, 
$D$ is the distance from the star-forming galaxy to the target halo, 
$R_{\rm vir}$ is a virial radius of the target halo, 
$\alpha_{\rm B}$ is the case B recombination coefficient,
$n_{\rm H}$ and $n_{\rm H, IGM}$ are the mean hydrogen number density in the halo and
the intergalactic medium, respectively. 
$n_{\rm H}$ and $n_{\rm H, IGM}$ just depend on redshift, 
$n_{\rm H} = \Delta_{\rm c} \times n_{\rm H, IGM}^{z=0} (1+z)^{3}$, 
where $\Delta_{\rm c} \sim 200$ is the mean over density 
and $n_{\rm H, IGM}^{z=0} \sim 2.0 \times 10^{-7}~\rm cm^{-3}$. 
The second term on the right-hand side of the above equation
is the absorption of ionizing photons by the intervening IGM, 
and much smaller than the first term. 
Using above equation we estimate the critical distance for target haloes to be ionized by a neighbouring galaxy as 
\begin{equation}
\begin{split}
D_{\rm crit} < 73~{\rm kpc} \left( \frac{f_{\rm esc}^{\rm ion}}{0.1} \right)
 &\left(  \frac{\rm SFR}{10~\rm \Msun~yr^{-1}} \right) \\
 &\times \left(  \frac{\Mh}{10^{9}~\Msun} \right)^{-\frac{1}{3}}
\left(   \frac{1+z}{8} \right)^{-5}
\end{split}
\end{equation}

$f_{\rm esc}^{\rm ion}$ is the escape fraction of ionizing photons from the neighbouring galaxy
\citep{Yajima09, Yajima11, Yajima14c, Paardekooper15, Wise14}, 
$\Mh$ is the halo mass of the target.
Here we use the relation $\dot{N}_{\rm ion} = 0.93\times10^{53}~{\rm s^{-1}}~\rm \left(  \frac{SFR}{\Msunyr} \right)$ 
for the Salpeter IMF  \citep{Madau98}. 
For a Chabrier IMF, the number of ionizing photons will be a factor two higher.
Observed LBGs at $z \gtrsim 6$ have high SFRs $\gtrsim 10~\Msunyr$ \citep[e.g.,][]{Bouwens15}. Haloes residing near these bright galaxies within $D \lesssim 70~\rm kpc$ will be ionized.

\subsection{Angular momentum transport}
The formation of a massive POP-III cluster requires the efficient transport of gas towards the haloes potential minimum on a time scale short enough  $\sim 4~\rm Myr$ to avoid the onset of super-novae feedback from POP-III stars. 
One obstacle is the presence of angular momentum which will lead to the formation of a large scale disc, if conserved. 
In the absences of molecular hydrogen, gas in haloes with  virial temperature higher than $\sim 10^4$ K can start to collapse. 
Then, as the gas density increases, self-shielding against the ionizing radiation from neighbouring star forming galaxies will kick in, 
and the gas becomes fully/partially neutral. 
If the virial temperature is $\lesssim 10^{5}~\rm K$, the thermal energy is released by $\lya$ radiation cooling. 
$\lya$ photons escape from the halo after they are trapped for a while due to scattering. 
As a result, the photon number density becomes high. 
\citet{Yajima14d} showed the angular momentum is efficiently transported due to $\lya$ radiation drag. 
After virialization, the thermal energy is released over a dynamical time of the halo as follows.
\begin{equation}
\begin{split}
U \sim \frac{G\Mh^{2}f_{\rm gas}}{\Rvir}, ~~
\La \sim \frac{U f_{U, \rm Ly\alpha}}{\tdyn (\overline{\rho})}, 
\end{split}
\end{equation}
where $\Mh$ is halo mass, $\fgas$ is gas mass fraction to dark matter, $f_{U, \rm Ly\alpha}$ is the fraction of thermal energy released in form of $\lya$ radiation,
$\tdyn (\overline{\rho})$ is the dynamical of the halo at its mean density, $\overline{\rho} = \Delta_{\rm d} \times \rho_{0}(z)$, $\rho_{0}(z)$ is the mean cosmic matter density at the specific redshift. 
In this paper we set the halo mean over density as $\Delta_{\rm c}=200$, because our focus is the early Universe at $z \ge 6$  when the energy density in the Universe is dominated by the matter density and the background cosmology is similar to the Einstein-de-Sitter model. The stacked $\lya$ photon number considering photon trapping is estimated as,
\begin{equation}
\begin{split}
\Na = \frac{\La t_{\rm trap}}{\ea} 
= \frac{\La 15 \ft 2 \Rvir}{c \ea},
\end{split}
\end{equation}
where $\ea=10.2~\rm eV$ is the energy of a $\lya$ photon, 
$t_{\rm trap}$ is the trapping time, 
and $\ft \equiv \frac{t_{\rm trap}}{15t_{\rm cross}}$.
Adams (1973) calculated the trapping time, 
and showed it is almost constant at $\tau < 10^{5.5}$,
and increases with $\tau$, 
as follows:
\begin{equation}
\ft = \begin{cases}
 1  ~~&{\rm for}~\tau < 10^{5.5}\\
\left(  \frac{\tau}{10^{5.5}} \right)^{\frac{1}{3}} ~~&{\rm for}~\tau \ge 10^{5.5},
\end{cases}
\end{equation}
where $\tau$ is optical depth to $\lya$ at the line centre.
If we assume  uniform density just after virialization, 
the optical depth over the halo is estimated by $\tau = 3.3 \times 10^{-14} \overline{n}_{\rm H}(z) 2\Rvir$ \citep[e.g.,][]{Verhamme06},
where $\overline{n}_{\rm H}(z) = 2.0 \times 10^{-2}\left(\frac{\Delta_{\rm c}}{200} \right) \left( \frac{1+z}{8} \right)^{3}$. 
Then we can derive $\ft$ for the halo with $\tau \ge 10^{5.5}$,
\begin{equation}
\begin{split}
 \ft = 3.7 \left( \frac{\Mh}{10^{9}~\Msun} \right)^{\frac{1}{9}}  
\left( \frac{1+z}{8}  \right)^{\frac{2}{3}}
\end{split}
\end{equation}
We estimate the $\lya$ photon number density in a halo as a function of the halo mass and redshift to be:
\begin{equation}
\begin{split}
\rhoa &= \frac{3\Na}{4 \pi \Rvir^{3}} 
= \frac{90 \ft G \Mh^{2} \fgas f_{U, \rm Ly\alpha}}{4 \pi c \Rvir^{3} \tdyn \ea} \\
&= 1.9 \times 10^{-4} \left( \frac{\Mh}{10^{9}~\Msun}  \right)^{\frac{10}{9}} \left( \frac{1+z}{8} \right)^{\frac{31}{6}}  \\
&~~~~~~~~~~~~~~~~~~~~~~~~~~ \times \left(  \frac{\fgas}{0.16} \right) f_{U, \rm Ly\alpha}~{\rm cm^{-3}}.
\end{split}
\end{equation}
A high background of $\lya$ radiation will exert a drag on   H{\sc i} gas clouds moving within it,
because of the anisotropic radiation field due to the doppler shift and beaming effects. 
\citet{Yajima14d} estimate the drag force for a hydrogen test particle in the residual ionized bubble around a POP-III star, 
and provide the following fitting function, 
\begin{equation}
\Fdrag  = \begin{cases}
0.4 \times 10^{-32}~{\rm erg~cm^{-1}} \left( \frac{v}{1 ~\rm km~s^{-1}}\right) & \left( \frac{\rhoa}{2.5 \times 10^{-3}~\rm cm^{-3}} \right)  \\
~&{\rm for}~ v < 17~{\rm km~s^{-1}} \\
0.4 \times 10^{-31}~{\rm erg~cm^{-1}} \left( \frac{v}{20 ~\rm km~s^{-1}}\right) & \left( \frac{\rhoa}{2.5 \times 10^{-3}~\rm cm^{-3}} \right)  \\
~&{\rm for}~ v > 17~{\rm km~s^{-1}}
\end{cases}
\end{equation}

\begin{figure}
\begin{center}
\includegraphics[scale=0.48]{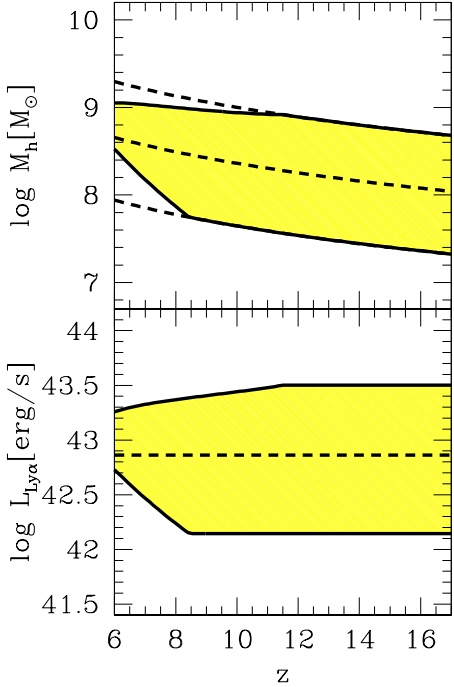}
\caption{
{\it Top panel:} Halo mass range where angular momentum transport efficiently occurs due to $\lya$ radiation drag. 
Dash lines represent virial temperatures of  $1\times10^{4}, ~3\times10^{4}~\rm K$ and $8 \times 10^{4}~\rm K$, respectively. 
{\it Middle panel:} $\lya$ luminosity expected in the halo mass and redshift. Dash line shows the $\lya$ luminosity for halo with the virial temperatures of  $3\times10^{4}$ K.
}
\label{fig:mcrit}
\end{center}
\end{figure}

Using the above equation, we consider the loss of angular momentum of gas to the  $\lya$ radiation background.
Here we assume an isothermal density profile for the dark matter halo, for which  the constant circular velocity is given by $v_{\rm c} \sim \sqrt(G\Mh/\Rvir)$ and estimate the time scale until $v_{\rm c}$ becomes small as:
\begin{equation}
\begin{split}
\tang &=  \frac{m_{\rm H} v_{\rm c}}{|\Fdrag (v_{\rm c})|} \\
&= 1.0 \times 10^{8}~{\rm yr} \left( \frac{\Mh}{10^{9}~\Msun} \right)^{-\frac{4}{9}}
\left( \frac{1+z}{8} \right)^{-\frac{25}{6}} \\
&~~~~~~~~~~~~~~~~~~~~~~~~~~~~\times \left( \frac{\fgas}{0.16} \right)^{-1} f_{U, \rm Ly\alpha}^{-1}.
\end{split}
\end{equation}
In this work, we focus on haloes with $v_{\rm c} > 17~\rm km~s^{-1}$. 
If $\tang \lesssim \tdyn$, the galaxy is likely to form a compact gas cloud at the galactic centre instead of a  large sale disc. 
We estimate the critical halo mass by setting  $\tang = \tdyn$,
\begin{equation}
\Mcrit^{\rm drag} = 1.5 \times 10^{8}~ \Msun \left( \frac{1+z}{8} \right)^{-6} \left( \frac{\fgas}{0.16} \right)^{-\frac{9}{4}} f_{U, \rm Ly\alpha}^{-\frac{9}{4}}, 
\label{eq:mdrag}
\end{equation}
%
If a halo is more massive than $\Mcrit$,  collapse to a central gas cloud is very efficient and a massive POPIII starburst can take place.
As the halo mass and corresponding virial temperature increases, thermal energy will  be released via He line cooling and free-free emission, 
resulting in $f_{U, \lya} \ll 1$. This leads to an upper limit in the halo mass due to inefficient $\lya$ radiative cooling. 
The corresponding halo mass range is shown in Figure~\ref{fig:mcrit}. Gas in haloes with masses $10^{8} \lesssim \Mh/\Msun \lesssim 10^{9}$ can loose its angular momentum efficiently.
We consider other lower and upper limits corresponding to $\Tvir=10^{4}~\rm K$ and $8\times10^{4}~\rm K$. 
The temperature of ionized gas depends on the flux and shape of the SED of the source. 
If the source is an AGN emitting hard X-rays, it can doubly ionize helium and increase the temperature to more than $5 \times 10^{4}~\rm K$, however it is difficult to achieve gas hotter than $8\times10^{4}~\rm K$ with reasonable spectral indices. Hence we set  $8 \times 10^{4}~\rm K$ as the upper limit. 
For a source with Saltpeter-like IMF and without AGN, the temperature of the ionized gas is usually less than $\sim 3\times10^{4}~\rm K$, corresponding  
 halo mass of 
$\Mh=3.7\times10^{8}~\Msun \left( \frac{1+z}{8}\right)^{3/2}$.

In the next subsection we will estimate the resulting star formation for such a collapse.
In this work, we assume that haloes can collapse if the virial temperature becomes higher than that of ionized gas. 
This is because the mean gas density in haloes at $z \gtrsim 6$ is $n_{\rm H} \sim 10^{-2}~\rm cm^{-3}$,  marginally self-shielding \citep[e.g.,][]{Yajima12h}. 
Hence, once the gas temperature is smaller than $T_{\rm vir}$, gas collapses  and becomes neutral due to the self-shielding, and then quickly cools down via $\lya$ emission, resulting in star formation. 
Note that, the estimation of the radiation drag and the critical halo mass assumes primordial gas.
For galaxies that are dust enriched, the $\lya$ photon density does not increase due to  dust absorption and in these cases you will form discs, even in haloes more massive than $\Mcrit^{\rm drag}$.
\citet{Hartwig16} also indicated that most haloes could be metal polluted at high redshift, resulting in the suppression of massive POPIII starbursts at $z \lesssim 8$. 

\subsection{Star formation rate}
Gas with negligible angular momentum collapse to the centre of a halo over a halo dynamical timescale and converts there into stars efficiently. 
We can write the empirical star formation law as:

\begin{equation}
\begin{split}
& \Msf  \sim \dot{M}_{\rm acc} = \fang \frac{\Mh f_{\rm gas}}{\tdyn(\overline{\rho})}  \\
&= 6.6\times 10^{-1}~{\rm \Msun~yr^{-1}} \fang \left( \frac{\Mh}{10^{9}~\Msun} \right)
\left( \frac{f_{\rm gas}}{0.16} \right) \left( \frac{1+z}{8} \right)^{\frac{3}{2}},
\end{split}
\label{eq:sfr}
\end{equation}
where $\fang = {\rm min}[\frac{\tdyn}{\tang},1]$.  Now we focus on the situation $\tang < \tdyn$, which leads to a POPIII starburst due to efficient angular momentum transport. 
On the other hand, if angular momentum is not transported efficiently, i.e., $\tang \gg \tdyn$,
a rotationally supported disk forms with $R_{\rm disk} \sim \lambda \Rvir$ \citep[e.g.,][]{Mo98},
where $\lambda$ is the spin parameter. 
In the latter case star formation will proceed quietly following a Kennicutt-Schmidt law, 
resulting in $\esf \lesssim 1$, and it is unlikely that this will produce an observable POPIII-dominated galaxies.

The life time of POPIII stars is $\sim 4~\rm Myr$ \citep{Schaerer02}. After the initial starburst, POPIII supernovae produce metals and evacuate most gas from a halo. 
Therefore, the maximum stellar mass is estimated by multiplying the life time to the star formation rate of Equation~(\ref{eq:sfr}) for fixed $\fgas$,
\begin{equation}
M_{*} = 2.6\times10^{6} ~\Msun~\fang \left( \frac{\Mh}{10^{9}~\Msun}\right)
\left( \frac{\fgas}{0.16}\right) \left( \frac{1+z}{8} \right)^{\frac{3}{2}}
\label{eq:stmass}
\end{equation}
Most of the $\lya$ emission comes from recombinations in gas previously ionized
by UV radiation from POPIII stars. Thus the $\lya$ luminosity is proportional to the ionizing photon emissivity of stars $\dot{N}_{\rm ion}$, and we estimate it as:
\begin{equation}
\begin{split}
\La &= 0.68 (1 - f_{\rm esc}^{\rm ion}) \ea \dot{N}_{\rm ion} \\
    &\sim 0.7 \times 10^{44}~{\rm erg~s^{-1}} \left( \frac{M_{*}}{10^{7}~\Msun}  \right)
\end{split}
\label{eq:lum}
\end{equation}
where $f_{\rm esc}^{\rm ion}$ is escape fraction of ionizing photons.
$f_{\rm esc}^{\rm ion}$ is typically $\lesssim 0.1$ \citep[e.g.,][]{Yajima11}. 
The effective temperature of POPIII stars is $\sim 10^{5}~\rm K$, and the mass dependence is very weak \citep{Bromm01a}.
Hence,  the total ionizing photon emissivity is proportional to the total stellar mass independent to the detailed initial mass function \citep[e.g.,][]{Susa14}. 
We estimate the total ionizing photon emissivity by using that of a $40~\Msun$ POP III star calculated in \citet{Schaerer02}, resulting in the relation, $N_{\rm ion}=6.2\times 10^{47} \left( \frac{M_{*}}{\Msun} \right)$. 
Hence, combining Equation~(\ref{eq:stmass}) and (\ref{eq:lum}) shows the relation between $\lya$ luminosity and halo mass at a specific redshift,
\begin{equation}
\La = 2.0 \times 10^{43}~{\rm erg~s^{-1}}~\fang \left( \frac{\Mh}{10^{9}~\Msun}\right)
\left( \frac{\fgas}{0.16}\right) \left( \frac{1+z}{8} \right)^{\frac{3}{2}}
\label{eq:lum2}
\end{equation}
The range of $\lya$ luminosity is shown in the lower panel of Figure~\ref{fig:mcrit}. 
Our model indicates that POPIII-dominated galaxies have $\lya$ luminosity of $\sim 1\times10^{42} - 3\times10^{43}~\rm erg~s^{-1}$. 
The $\lya$ luminosity of CR7 is much higher than the upper limit predicted by of our model. 
Therefore,  we argue that CR7 is not a  POPIII star cluster.
This is in agreement with recent observational data from Spitzer/IRAC that indicates  [OIII] emission from CR7 \citep{Bowler16}, indicating an enriched population of stars.  

The predicted $\lya$ luminosity range is detectable even with current telescopes, suggesting that LAEs, which have high EW and are in the luminosity range near star-forming galaxies or AGNs, can be POPIII-dominated galaxies. 
On the other hand, the He{\sc ii} $1640~\rm \AA$ line flux can be lower than the $\lya$ line by $\sim 1-3$ orders    depending on the IMF of POPIII stars \citep{Schaerer02}
and hence its flux is difficult to detected with current telescopes.
Spectroscopy with JWST will be able to detect the He{\sc ii} lines within a reasonable integration time as a signature of POPIII-dominated galaxies. 
The threshold stellar mass with integration times of $\lesssim 10$ hours is $\Mstar \sim 10^{6}~\Msun$ assuming a top-heavy POPIII IMF with $\gtrsim 100~\Msun$.




\subsection{Outflow of gas due to Ly$\alpha$ radiation pressure}

As POPIII stars enter the ZAMS (zero age main sequence), 
they start to exert feedback on the interstellar medium (ISM) and may cause galactic outflows. 
Due to the trapping of $\lya$ radiation in a halo, $\lya$ radiation pressure can be higher than
the thermal pressure of ionized gas or radiation pressure by H{\sc i} photo-absorption or Thompson scattering
\citep{Yajima14a, Smith16}. 
The momentum equation is then approximately given by,
\begin{equation}
\Mh f_{\rm gas}  \frac{dV_{\rm out}}{dt} = \frac{15 \ft \La}{c}  - \frac{G\Mh^{2}\fgas}{\Rvir^{2}}.
\end{equation}
If we substitute the expression for the $\lya$ luminosity from Equation~(\ref{eq:lum}) into the first term on the right hand side, we find that 
the radiation pressure is higher than the gravitational force for haloes with 
$\Mh < 3.5 \times 10^{13}~\Msun~\ft^{3} \left(  \frac{1+z}{8} \right)^{-3/2}$, i.e., the radiation pressure can drive an outflow. 
For the halo mass range with  POPIII starbursts ($\Mh \sim 10^{8}-10^{9}~\Msun$), 
the gravitational force is negligible.  We estimate the outflow velocity considering only the $\lya$ radiation pressure 
by integrating the above equation over 4 Myr as
\begin{equation}
\Voutmax=39.5~{\rm km~s^{-1}} ~ \ft \left(  \frac{1+z}{8} \right)^{\frac{3}{2}}.
\end{equation}
%
The velocity is smaller than the escape velocity for haloes in  the mass range, 
$\Mh > 1.8\times10^{7}~\Msun ~\ft^{3} \left( \frac{1+z}{8} \right)^{3}$. 
Therefore, $\lya$ radiation pressure alone is not enough to evacuating gas from haloes.
POPIII supernovae then take over driving the wind. 
The gas outflow leaves its signature in the $\lya$ line profile
as an asymmetric shape with red wing  \citep[e.g.,][]{Yajima12b}. 
From the relation $\ft = 2.2 \left( \frac{N_{\rm HI}}{10^{20}~\rm cm^{-2}} \right)^{1/3}$ and the outflow velocity, 
the $\lya$ line profile can be calculated and compared to  high-dispersion spectra.
 
In this work, the outflow velocity does not depend on halo mass if $\ft$ is constant. 
However, \citet{Smith16b} showed that 
the outflow velocity weakly increases with halo mass due to longer photon trapping times. 
In addition, we here assume $\lya$ radiation feedback acts uniformly on the gas, 
while \citet{Smith16b} show that in the early phase the inner gas shells alone are affected by $\lya$ radiation pressure,
resulting in higher outflow velocity of the inner gas shells.

\section{Summary}
\label{sec:summary}
Massive population III (POPIII) starbursts are a viable way of forming POPIII star dominated proto-galaxies.
In order for a massive POPIII starburst to occur, star formation in mini-haloes needs to be suppressed due to photo-ionization from neighbouring galaxies.
$\lya$ radiation drag is efficient in removing angular momentum from collapsing gas, however, it is counter-balanced by $\lya$ radiation pressure.  
As a result, our model shows that POPIII-dominated galaxies are hosted by haloes of $\Mh \sim 10^{8} - 10^{9}~\Msun$, 
and  have $\lya$ luminosities of $\La \sim 1\times10^{42} - 3 \times 10^{43}~\rm erg~s^{-1}$ and outflowing gas with velocity $V_{\rm out} \sim 40~\rm km~s^{-1}$.
The maximum $\lya$ luminosity is much fainter than that observed by the POPIII galaxy candidate CR7 (Sobral et al. 2015), suggesting that CR7 is not a POPIII galaxy. 
Our predicted luminosity and wind velocity suggest that future spectroscopic survey with e.g. JWST and TMT will be able to detect POPIII dominated proto-galaxies.

%
%
\section*{Acknowledgments}
We are grateful to Yuexing Li and Kazuyuki Sugimura for valuable discussion.
We thank the anonymous referee for useful comments.
This study is supported in part by JSPS KAKENHI Grant Number 15H06022 (HY).

%
%
\bibliographystyle{mn}


\label{lastpage}

\end{document}